\documentclass[a4,12pt]{article}
\makeatother
\usepackage{indentfirst}
\usepackage[dvipdfm]{graphicx}
\usepackage{amsmath}
\usepackage{ascmac}
\usepackage{amssymb}
\usepackage{fancybox}
\usepackage[stable]{footmisc}
\newcommand{\beq}{\begin{eqnarray}}
\newcommand{\eeq}{\end{eqnarray}}

\newcommand{\bs}{\boldsymbol}
\title{
{\bf ``Quark Confinement'' and Evolution of \\ 
Covariant Hadron-Classification Scheme
}}
\author{Shin Ishida\footnote{Senior Research Fellow}
\vspace{1em}\\
{\it Research Institute of Science and Technology, Nihon University, } \\
{\it Tokyo 101-8308, Japan}
\date{\empty}
\vspace{1em}\\
E-mail: {isida@phys.cst.nihon-u.ac.jp}
        }
\begin{document}
\maketitle
\vspace{-2em}
\abstract{
The extension of Non-Relativistic Classification scheme for Composite 
Hadrons to Covariant one is one of the most important problem in the 
hadron spectroscopy. 
It seems us recently 
the Covar. Classif. scheme entering into an evolutionary stage, 
by taking seriously Quark-Confinement into account. 

In section \ref{sec1}, firstly we give a brief history of our way of 
the extension on the kinematical framework, that is, from N.R. scheme
one, $SU(2)_{\sigma}\otimes O(3)_{L}$ (aside from the flavor freedom), to 
Covar. scheme, $\widetilde{U}(4)_{DS,\mathfrak{m}}\otimes 
O(2)_{\boldsymbol{r} \perp \boldsymbol{v}}$ 
(the former is a tensor-space of Dirac-spinor embedded with a static 
spin-symmetry $SU(2)_{\mathfrak{m}}$, ${\mathfrak{m}}$ representing 
a new mass-reversal symmetry reflecting the physical situation of 
Q.C.; while the latter is a space of 2-dimensional internal 
spatial-vector $\boldsymbol{r}$, being orthogonal to the boost velocity 
$\boldsymbol{v}$, embedded in the Lorentz-space $O(3,1)_{\rm Lor.}$). 
Secondly we describe a role of the chirality symmetry in 
Composite Hadrons, which is valid through light to heavy quark system. 
It is a symmetry of QCD/Standard Gauge Model, of which importance 
in hadron spectroscopy has been overlooked for many years. 

In section \ref{sec2}, propertime $\tau$ quantum mechanics for 
multi-body confined quark system and quantization of Comp. 
Hadron field is developed. The similar to conventional procedures 
are performed, but all in Galilean 
inertial frame (with $\boldsymbol{v}=\rm const$); starting from an 
application of variational method to a classical action of the 
relevant confined system, where quarks have Pauli-type 
$SU(2)_{\sigma}$-intrinsic spin and also $SU(2)_{\mathfrak{m}}$-mass spin. 
A notable feature of the $\tau$-Q.M. is, it is concerned only 
future-development : which induces application of the 
crossing rule for ``Negative-Energy Problem''. 
This rule is conventionally supposed ad hoc. 
The $\tau$-Q.M. also induces Existence of the chiral-quark, 
with $J^P=(1/2)^{-}$ (, in addition to 
the normal quark with $J^P=(1/2)^{+}$) which 
is considered to be an origin of new 
exotics, mysterious from N.R. scheme. 

In section \ref{sec3}, it is summarized somewhat new framework 
of hadron spectroscopy guided by $\tau$-Q.M.: 
Especially, ``Regge Trajectories'', 
in $q\bar{q}$ meson system, are given by 
Mass-Squared vs. $\hat{N}(n,l)$; where $M^2=M_{0}^2+N \Omega$ 
 (
 $N=2n+l$), 
Intrinsic spin of hadrons 
$\boldsymbol{J}=\boldsymbol{S}$ {\it comes from only quark-spin}, 
and {\it Orbital $\boldsymbol{l}$ 
contributes only to N}. 

In section \ref{sec4}, the evolved Cov. Class. scheme 
is applied to phenomenology of bottomonium, 
of which experiment obtained a great 
progress recently. As a result it has been shown the 
{\it several remarkable facts}, including the 
{\it seemingly-strange ones from N.R. framework, are possible to 
be clearly explained/interpreted}. 
 }
 \newpage
\section{Introduction}
\label{sec1}
\begin{figure}[htbp]
\centering
  \fbox{
 \includegraphics[width=12cm, bb=0 0 750 550]{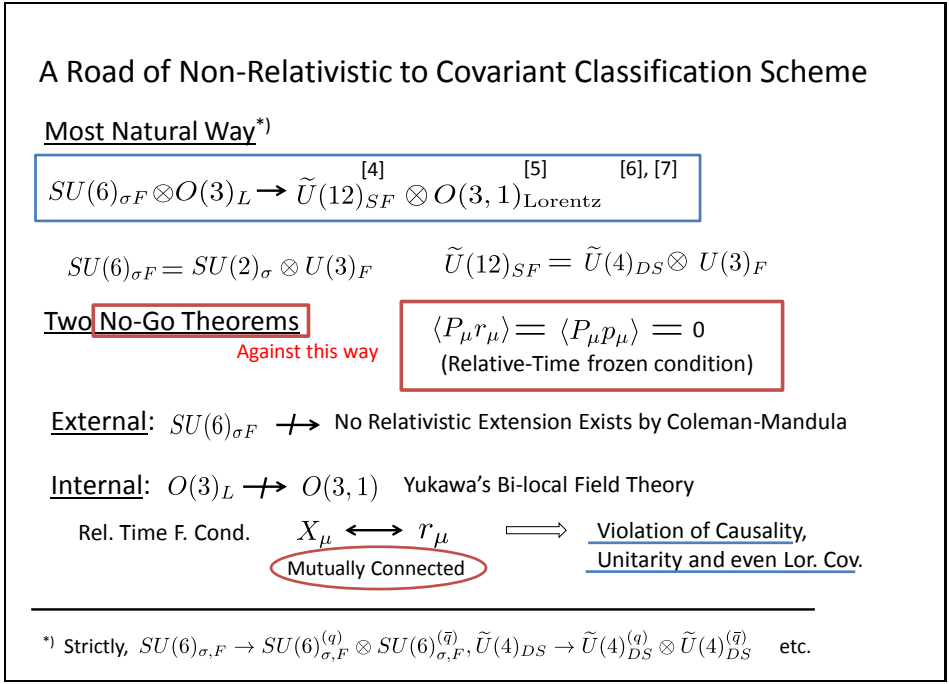}
  }
   \caption{``No-Go'' Theorems against Covariant Extension of 
   Non-Relativistic Spectroscopy. }
   \label{fig1}
 \end{figure}
\begin{figure}[htbp]
\centering
  \fbox{
  \includegraphics[width=12cm, bb=0 0 750 550]{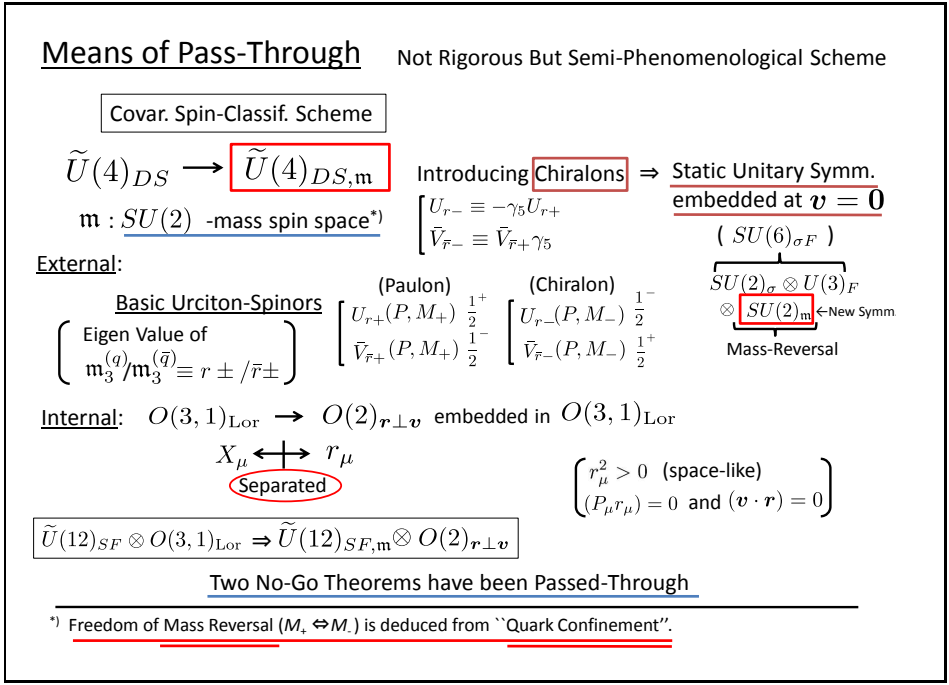}
  }
   \caption{Phenomenological Means of Pass-Through Difficulties. }
   \label{fig2}
 \end{figure}
\subsection{A Road of Non-Relativistic to Covariant Classification Scheme
\cite{Ref1,Ref2,Ref3}}
\label{sec1.1}

\indent Most natural way{\cite{Ref4,Ref5,Ref5-2,Ref5-3}} of extension is to extend separately both of external and internal parts of the kinematical framework(, see Fig.{\ref{fig1}}).
However, this way had been seen to be closed by two 
No-Go theorems.{\cite{Ref6}} 
The external one by the rigorous mathematical one; while the internal one as result of detailed investigation by Yukawa-School 
researchers.{\cite{Ref7}}
The origin of internal No-Go theorem is considered to come from the 
close connection between the C.M. coordinate $X_{\mu}$ and 
the internal coordinate $r_{\mu}$, 
as is seen from an ad hoc subsidiary ``relative-time frozen'' condition. 
Then we have chosen a semi-phenomenological means 
of pass-through(, see Fig.{\ref{fig2}}). Concerning the intrinsic quark-spin, 
{\it a new freedom of $SU(2)$-mass spin} is supposed in addition to the 
$SU(2)$-$\sigma$ spin, which makes possible to apply the crossing rule 
to confined quarks, in conformity with {\it color-singlet condition of parent 
hadrons}; while concerning the undesirable connection of internal $r_{\mu}$ 
to C.M. $X_{\mu}$, it is separated by supposing $r_{\mu}$ to be spacelike and 
$\boldsymbol{r}$ to be {\it orthogonal}
{\footnote{
This implies that constituent quarks have lost a role of carrier of 
{\it orbital angular momentum}, 
as $\boldsymbol{l}=\boldsymbol{r} \times \boldsymbol{p} (\boldsymbol{v})
\stackrel{\boldsymbol{v}=0}{=}0$ in the observer frame
 ($\boldsymbol{v}=0$). (See, section \ref{sec2.2}) )
}}
to the boost velocity $\boldsymbol{v}$. 
\subsection{Chirality: Overlooked Symmetry in Hadron Spectroscopy}
\label{sec1.2}

\begin{figure}[htbp]
\centering
  \fbox{
  \includegraphics[width=12cm, bb=0 0 750 550]{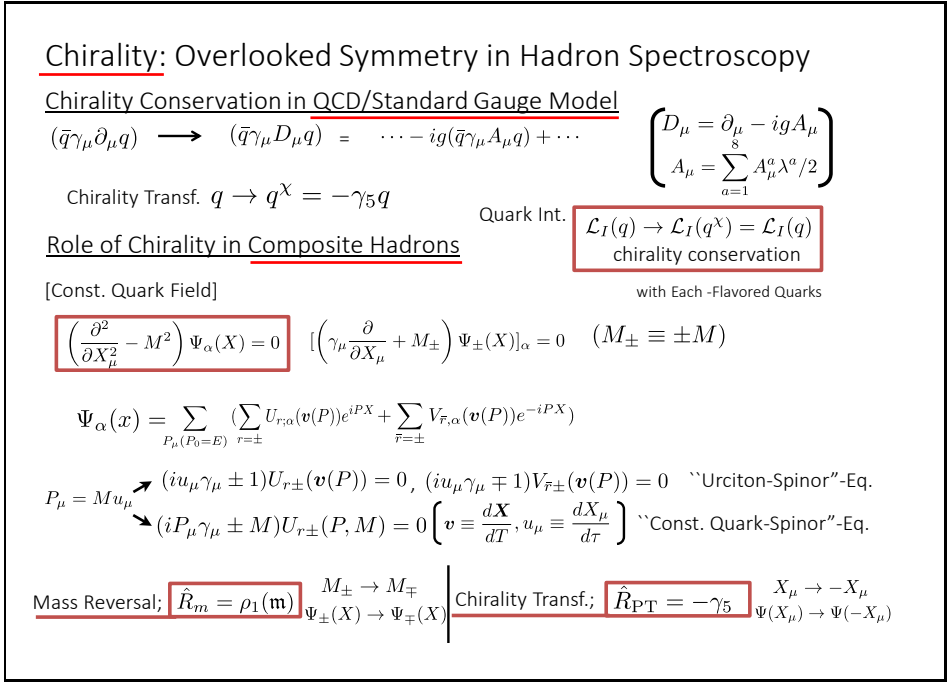}
  }
   \caption{Chirality Conservation in Standard Gauge Model, and Mass 
   Reversal of Constituent Quarks. }
   \label{fig3}
 \end{figure}

Chirality is(, see Fig.~{\ref{fig3}}), 
\begin{figure}[htbp]
\centering
  \fbox{
  \includegraphics[width=12cm, bb=0 0 750 550]{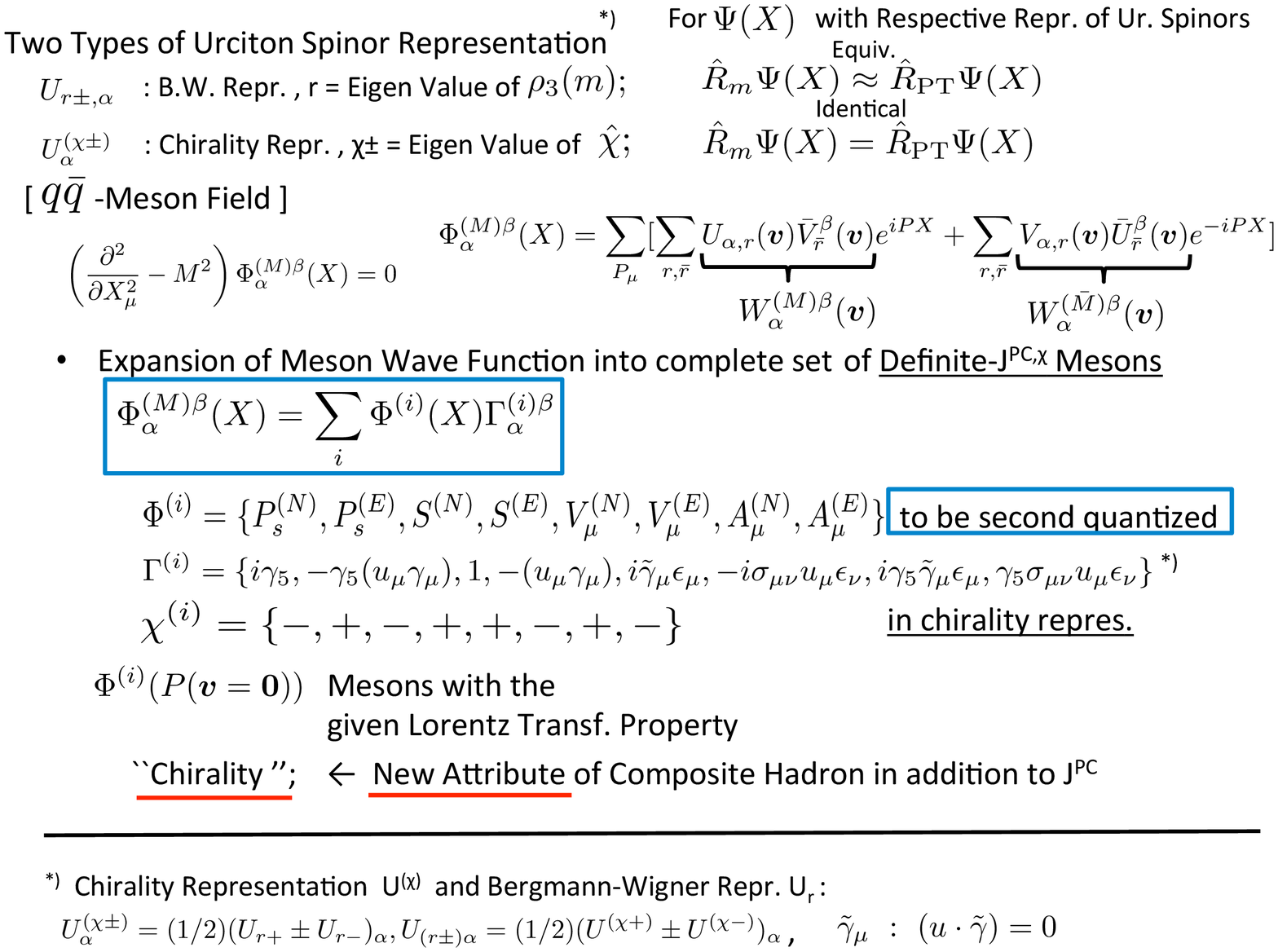}
  }
   \caption{New Attribute-$\chi \cdot$Structure of Composite Hadrons, and 
   Two Representations of Urciton-Quark Spinor. }
   \label{fig4}
 \end{figure}
an important symmetry 
conserved through all gauge-interaction with each-flavored quarks. 
This property of chirality seems us that it deserves to be an attribute 
of, somewhat {\it elementary entity, } Composite Hadrons, consisting of 
{\it confined quarks}. 

Here (Fig.~\ref{fig3}) it may be worth to point out the two side-role of 
Dirac-spinor in the covariant classification 
$\widetilde{U}(4)_{DS,\mathfrak{m}}$-scheme: the one, as urciton spinors, 
which concerns the chirality{\cite{Ref8}} structure $\chi^{(i)}(\pm)$'s of 
parent hadrons, while the other, as constituent-quark spinors concerns the 
mass-reversal{\cite{Ref9}}{
\footnote{
Here it may be notable that the sign of mass, which had been meaningless 
for free Dirac particles, now plays an important role for {\it Confined Quarks}: 
and that Galilean Inertial Frame with definite 
Boost Velocity ($\bs{v}\neq \bs{0} $) 
for Isolated, multi-particle system seems to be well representing the physical 
situation of Quark Confinement. The frame with  ($\bs{v}\neq \bs{0} $), 
to be called the Particle Frame, while the one with ($\bs{v}= \bs{0} $) is 
called the Observer Frame. The formulas obtained in the former (latter) 
becomes Lorentz invariant (covariant). 
}} structure $r^{(i)}(\pm)$'s of parent hadrons. 

Further (Fig.~\ref{fig4}) there are two 
Representations of Urciton spinors(, see Fig.{\ref{fig4}}); 
the B.W. Repr. with definite sign of Mass term and Chirality 
Repr. with definite 
sign of $\chi (\pm)$. The local meson wave-function 
$\Phi_{\alpha}^{(M)\beta}(X)$ 
is expanded by complete set of 8$\cdot$Dirac-Matrices, 
of which respective amplitudes, 
$\Phi^{(i)}$'s with {\it definite} $J^{PC\chi}$ are 
to be second-quantized, as ``{\it elementary}'' entity.
\footnote{
It is to be noted that Urciton spinors in $B.W.$ Repr. appear as 
spin W.F. of mixed-states of the quantized pure-states with definite $\chi$(see, 
section \ref{sec4}). 
}
\section{Propertime Quantum Mechanics for Multi-Body Confined-Quark
System and Quantization of Composite-Hadron Fields}
\label{sec2}

Before going into details, is shown overview for constructing a 
Quantum Field Theory of Composite Particle. 
\begin{figure}[htbp]
\centering
  \fbox{
  \includegraphics[width=12cm, bb=0 0 750 550]{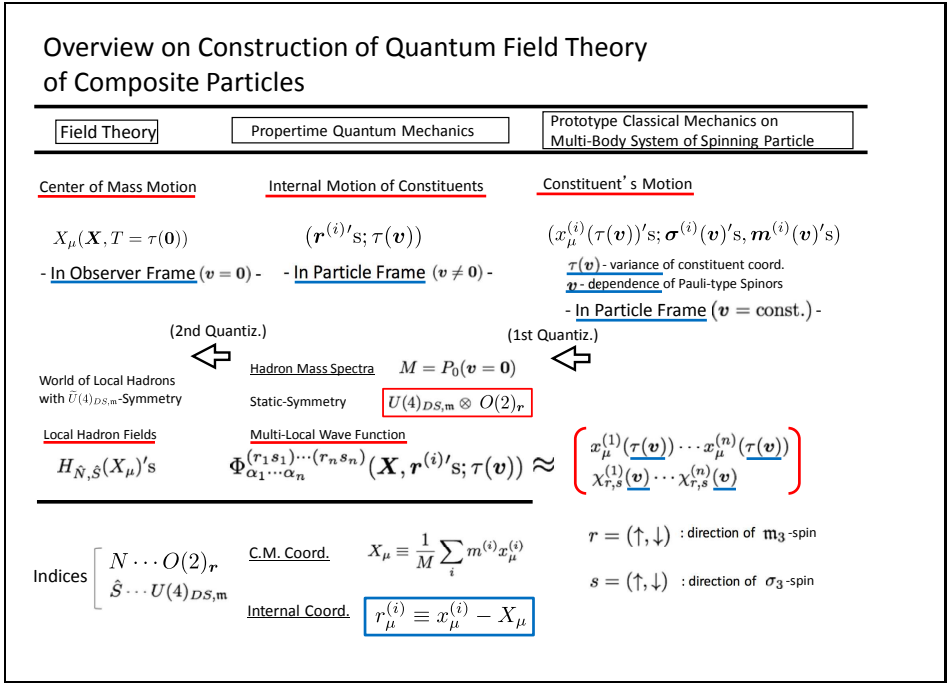}
  }
   \caption{Overview Bound for Composite-Hadron Field. }
   \label{fig5}
 \end{figure}
\subsection{Prototype Mechanics for Solitary Urciton-Quark Field
\footnote{Strictly this case is included as a part of General Formalism 
for multi Urciton-Quark system.}}
\label{sec2.1}
\begin{figure}[htbp]
\centering
  \fbox{
  \includegraphics[width=12cm, bb=0 0 750 550]{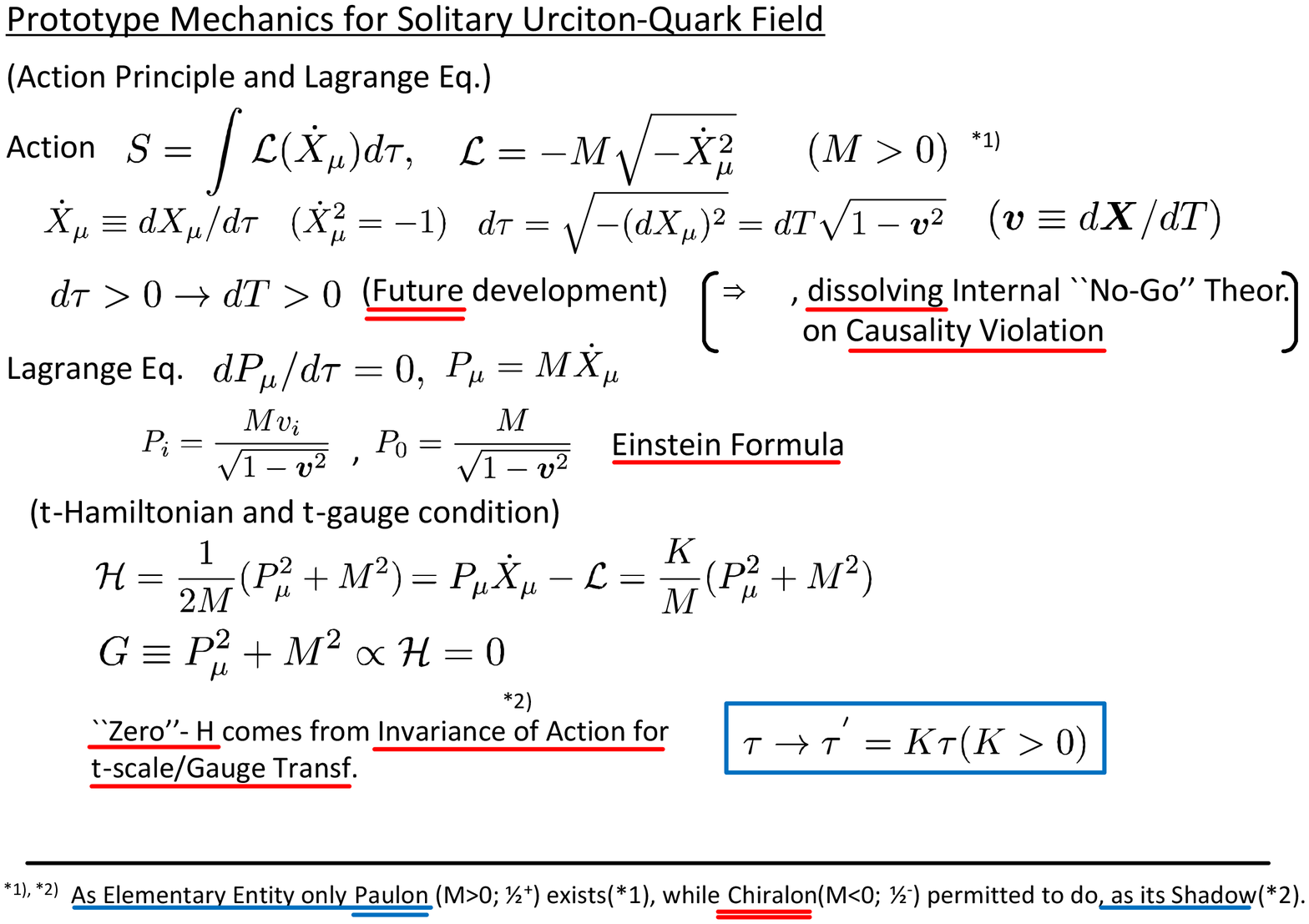}
  }
   \caption{Action Principle and $\tau$-Gauge Condition. }
   \label{fig6}
 \end{figure}
\begin{figure}[htbp]
\centering
  \fbox{
  \includegraphics[width=12cm, bb=0 0 750 550]{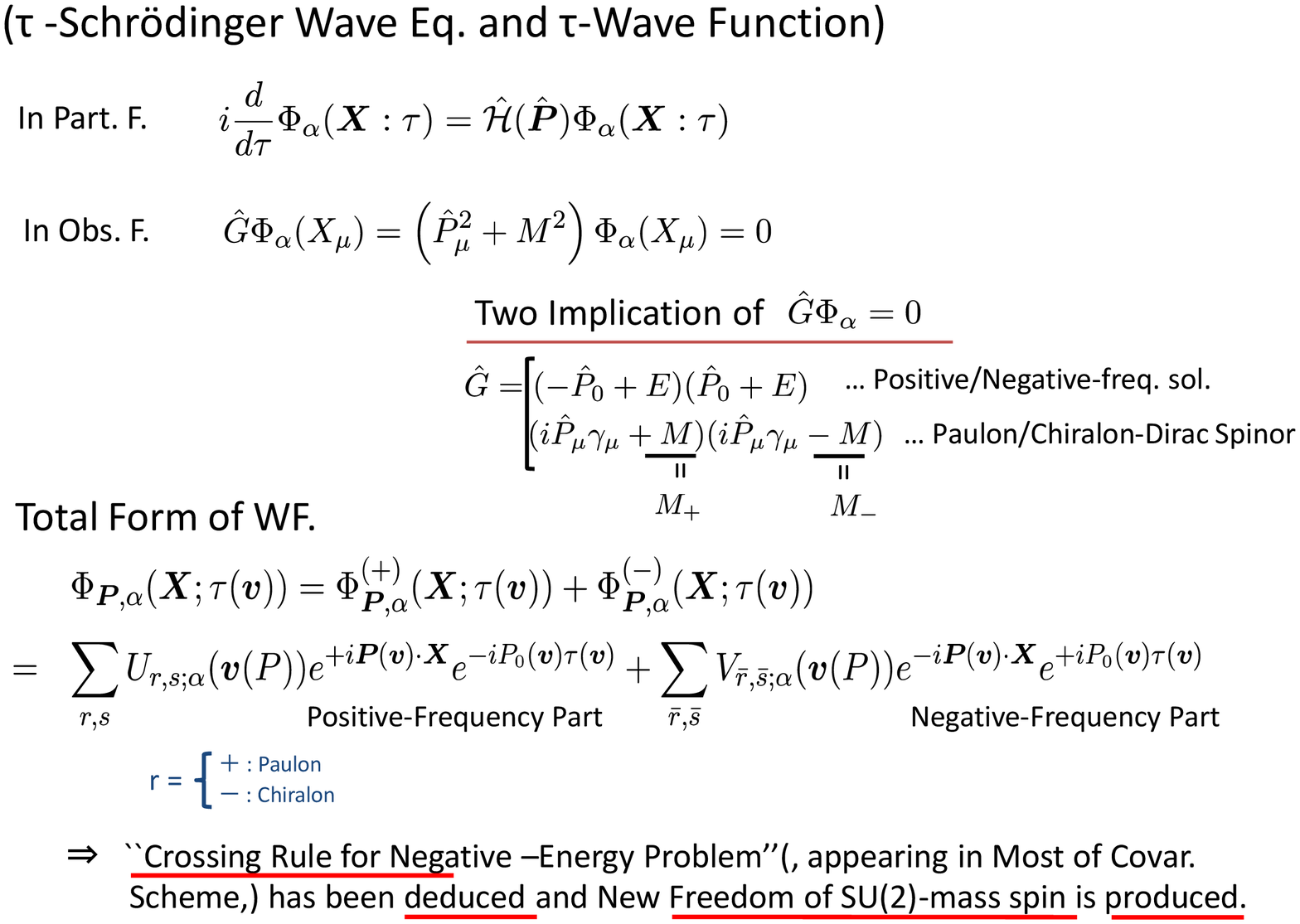}
  }
   \caption{$\tau$-Schr\"{o}dinger Wave-Equation and Existence of {\it Chiralon} as 
   {\it ``Shadow''} of Paulon. }
   \label{fig7}
 \end{figure}
The concrete formulas, obtained from application of the plan (Fig.~{\ref{fig5}}), 
in the ideal case of Constituent-Quark/Urciton Field is 
collected in Fig.~{\ref{fig6}} and 
~{\ref{fig7}}: From Fig.~{\ref{fig6}} we see that 
i) Einstein Formula on 4-momentum is included as 
Kinematics in our scheme; and 
ii) Mass-shell condition is deduced from the 
Invariance of action for $\tau$-Gauge 
transformation, which reflects that the $\tau$-scale of solitary/confined system 
is not observable. And from Fig.~{\ref{fig7}} we see that 
iii)~the $\tau$-gauge condition deduces on the one hand 
the $SU(2)_{\mathfrak{m}}$-mass space 
with the basic vectors{\footnote{
As Elementary Entity only Paulon ($M>0$; ${\frac{1}{2}}^{+}$) exits, 
while Chiralon($M<0$; ${\frac{1}{2}}^{-}$) permitted to do, as its Shadow. 
}} chiralon($J^{P}={\frac{1}{2}}^{-}$) in addition to 
Paulon($J^{P}={\frac{1}{2}}^{+}$); and 
on the other hand the Klein-Gordon Eq. as ``Prime'' one for Const. Quark Field, 
leading to the crossing rule{\footnote{
The reason of this deduction originates from that the propertime mechanics is concerned only 
{\it Future development}.
}} for Negative-Energy 
problem. 

\subsection{Prototype Mechanics for Multi-Urciton Quark Fields}
\label{sec2.2}
\begin{figure}[htbp]
\centering
  \fbox{
  \includegraphics[width=12cm, bb=0 0 750 550]{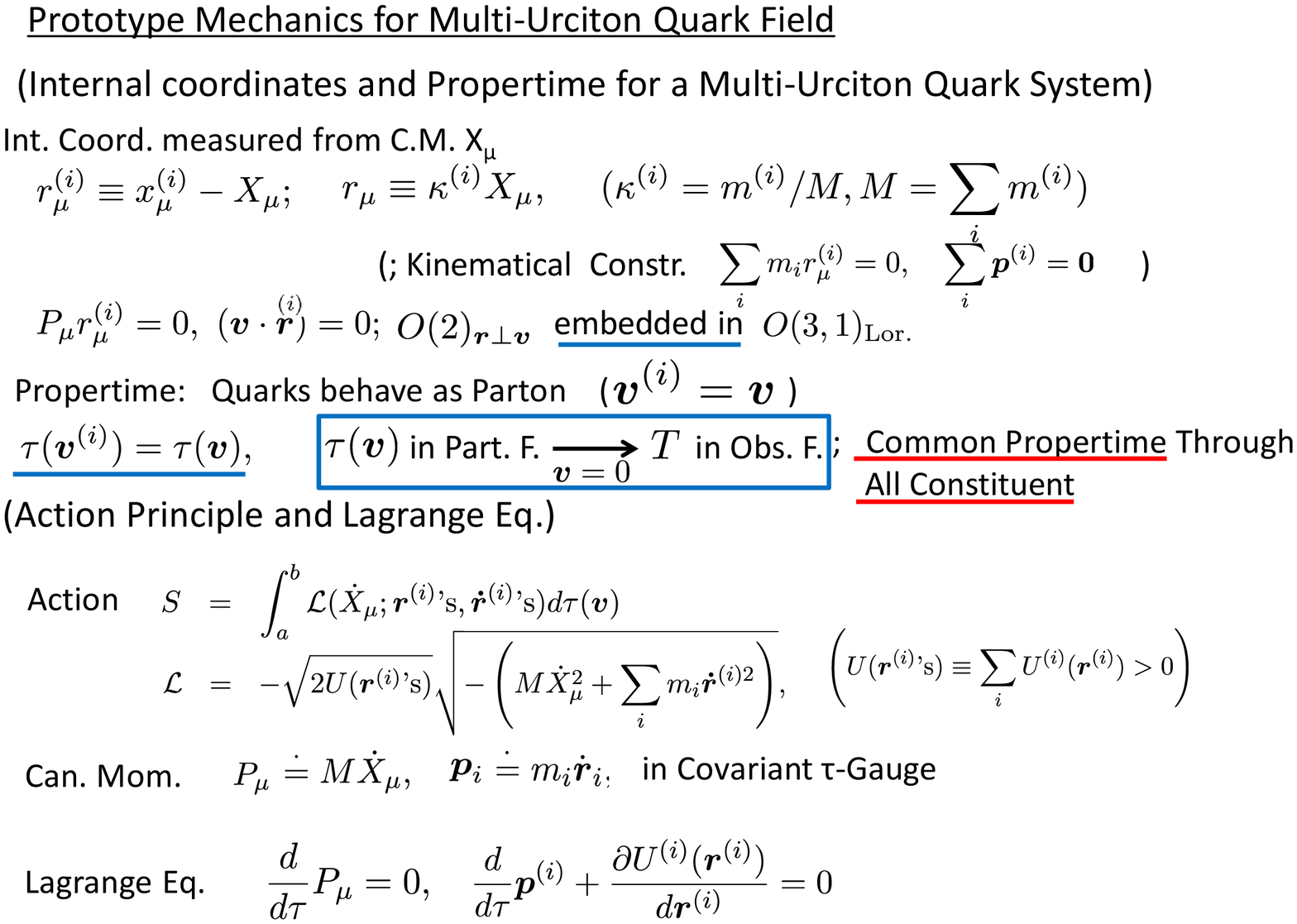}
  }
   \caption{Internal Coord., Propertime and Action Principle for 
   Multi-Particle System. }
   \label{fig8}
 \end{figure}
\begin{figure}[htbp]
\centering
  \fbox{
  \includegraphics[width=12cm, bb=0 0 750 550]{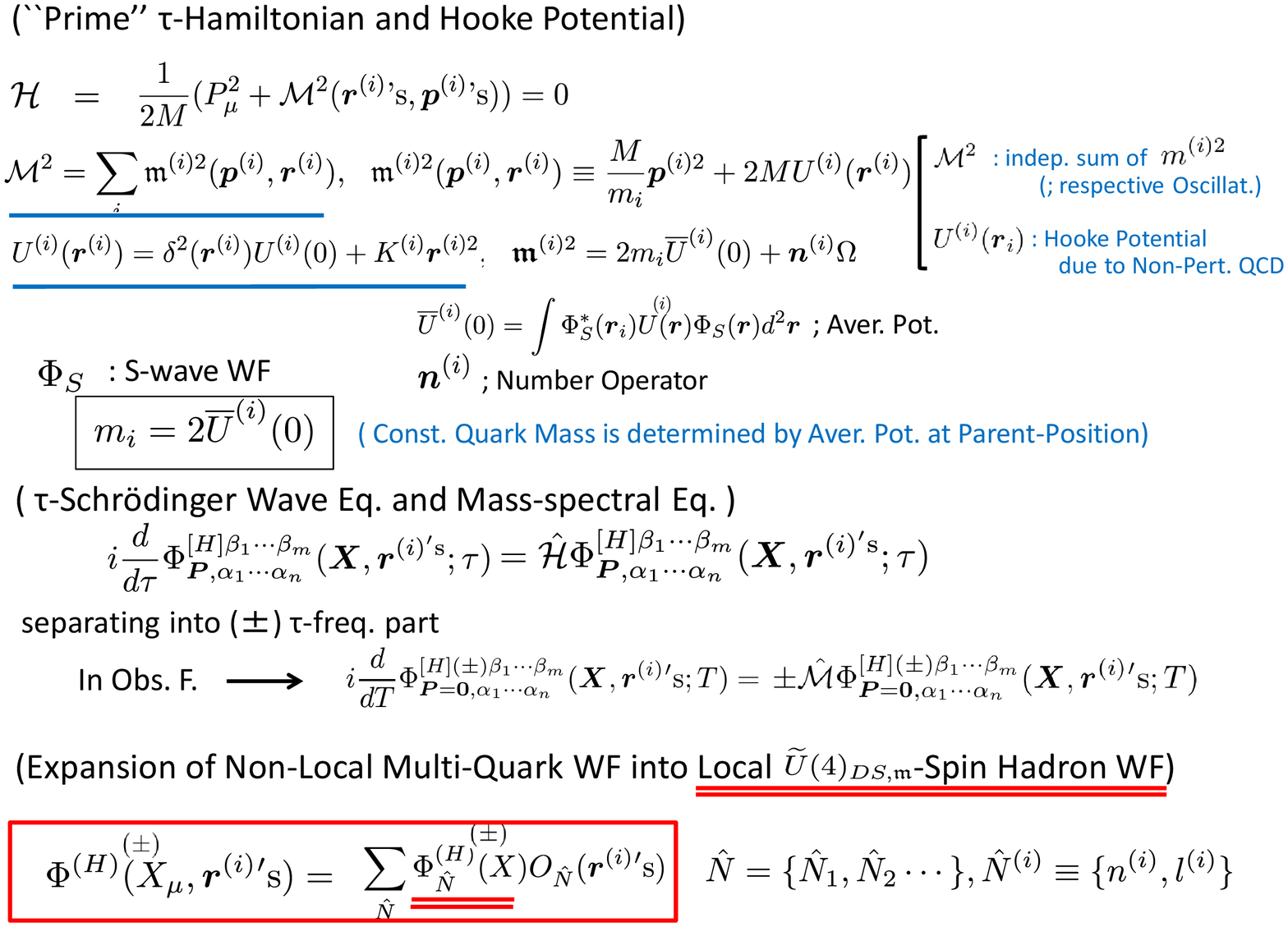}
  }
   \caption{Mass-Spectral Operator and 
   $\widetilde{U}(4)_{DS,\mathfrak{m}}$-Multiplet 
   of Local Hadron WF. }
   \label{fig9}
 \end{figure}
The concrete formulas, result of application of the plan (Fig.~{\ref{fig5}}) in the 
relevant case are collected in Figs. {\ref{fig8}} and {\ref{fig9}}: In Fig.~{\ref{fig8}} 
is also shown How to dissolve the undesirable connection between 
$X_{\mu}$ and $r_{\mu}$, both in $O(3,1)_{\rm Lor.}$ (in section \ref{sec1.1}),  
making the direct product $\{X_{\mu}$ in $O(3,1)_{\rm Lor.} \}$
$\otimes$
$\{\bs{r} \in  O(2)_{\bs{r} \perp \bs{v}}$, 
embedded in $O(3,1)_{\rm Lor.}\}$. 
Here it may be worth to note that the choice of internal coordinate 
$r_{\mu}^{(i)}$'s with the Kinematical Constraint (which, 
might be queer from the 
conventional view of composite model,) is quite natural in the Particle F., as 
an inertial frame. It is remarkable that there is No {\it Relative-Time Problem} 
in our scheme. 

In Fig.~{\ref{fig9}} is given, especially, (squared-)Mass Spectral Operator, which 
consists of independent sum of those, being respective Hooke-type oscillator on 
$\bs{r}^{(i)}$'s; and derived an interesting formula on the relation that 
the constituent-quark mass is given by the average value of respective potential 
$\bar{U}^{(i)}(0)$. Here also is given a Prescription of How to derive the Local Hadron 
WF to be Second Quantized from the Non-Local Multi-Quark WF. Here it may be 
instructive to note that all the results given in section \ref{sec2.1} are 
valid here (in section \ref{sec2.2}) as formulas concerning on C.M. coordinate $X_{\mu}$. 
Needless to mention, this comes from the separation between $X_{\mu}$ and 
$r_{\mu}$ explained in Fig.~{\ref{fig8}}. 
\section{Hadron Spectroscopy Guided by $\tau$-Quantum Mechanics}
\label{sec3}
\begin{figure}[htbp]
\centering
  \fbox{
  \includegraphics[width=12cm, bb=0 0 750 550]{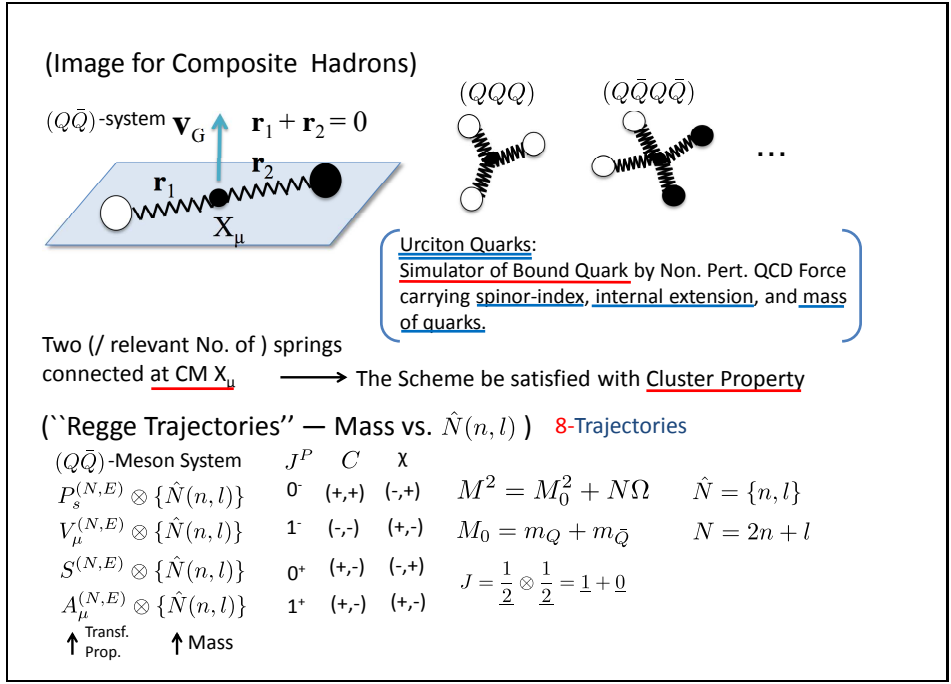}
  }
   \caption{Physical Image for Composite-Hadron System and Eight-Trajectories 
   of Quarkonium. }
   \label{fig10}
 \end{figure}
In Fig.~{\ref{fig10}} is summarized Essence of Renewed Covariant Classif. Scheme. 
It might be notable that such as the concrete picture for Composite-Hadron 
becomes from the view-point of Particle F.. In connection to this we should like to 
point out that Urciton Quarks are c-number Simulator of Bound Quarks by Non-Pert. 
QCD-Force, carrying spinor-index, internal extension; and mass of 
quarks; and that our scheme be satisfied{\cite{Ref9-2}} with Cluster Property. 
\section{Phenomenology of Bottomonium-System in the Evolved Classification-Scheme}
\label{sec4}
\begin{figure}[htbp]
\centering
  \fbox{
  \includegraphics[width=12cm, bb=0 0 750 550]{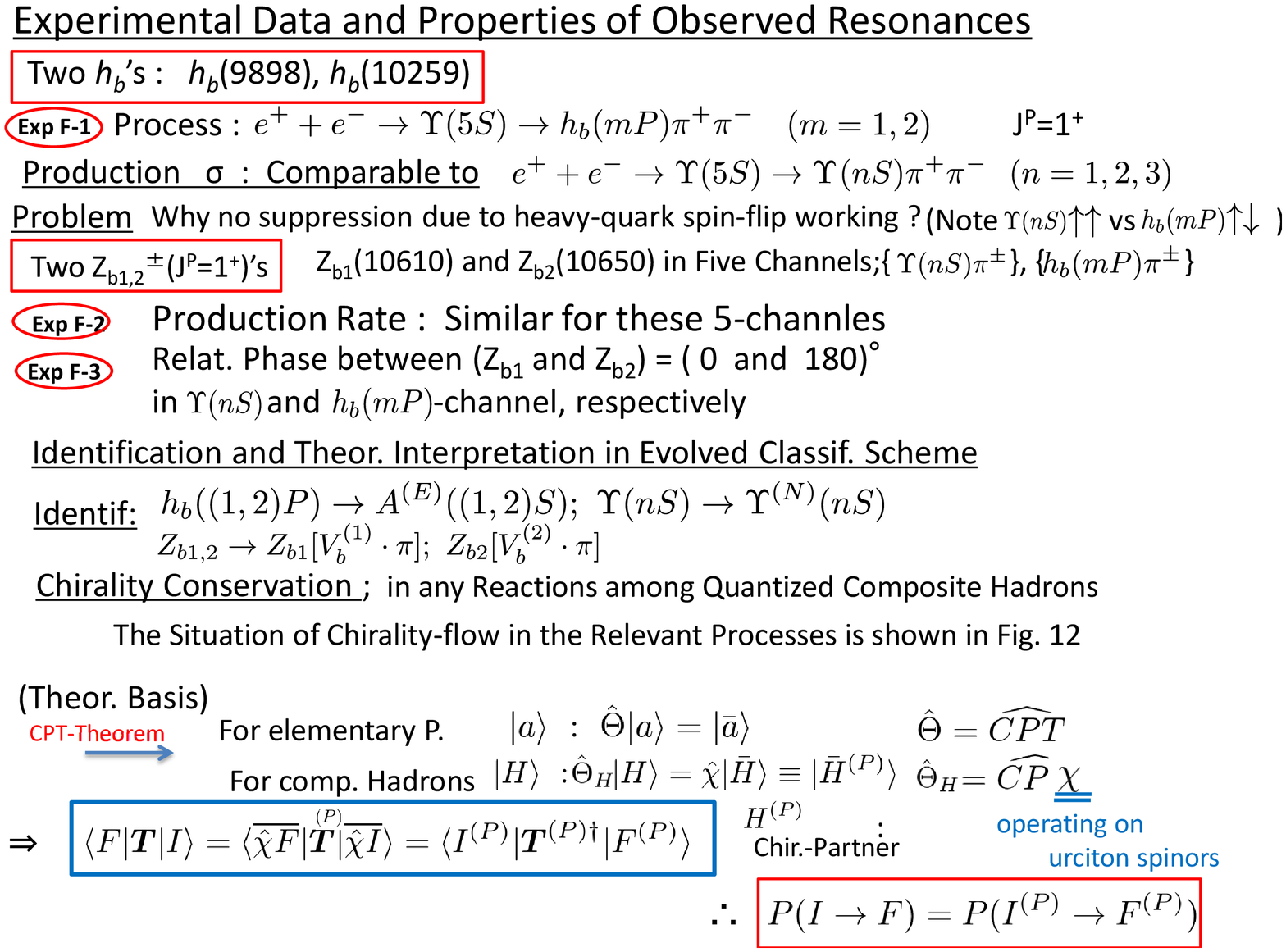}
  }
   \caption{Experimental Features of New Observed Resonances and 
   Its Interpretation in Evolved Classif. Scheme. }
   \label{fig11}
 \end{figure}
\begin{figure}[htbp]
\centering
  \fbox{
  \includegraphics[width=12cm, bb=0 0 750 550]{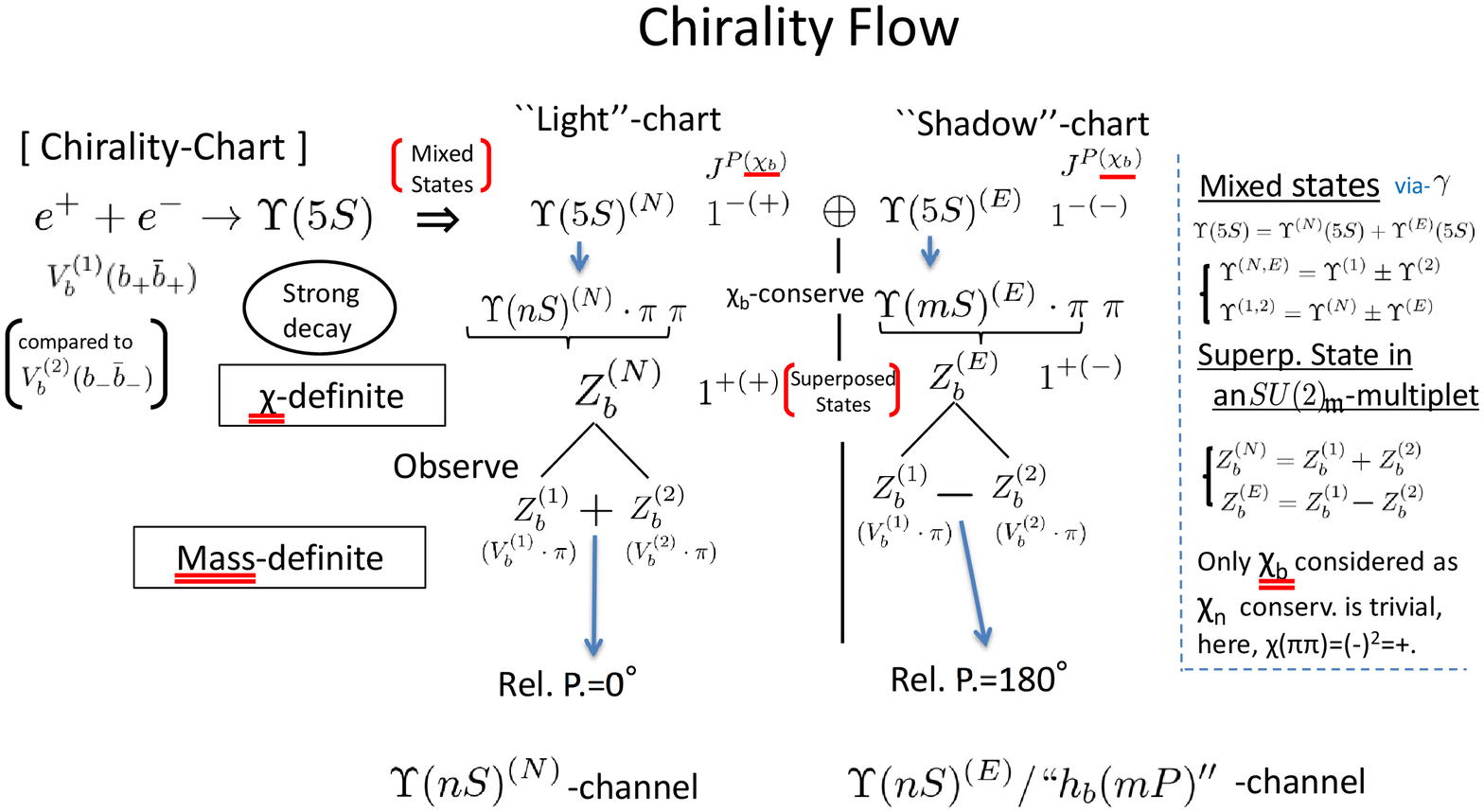}
  }
   \caption{Chirality-Flow Chart in Respective $\chi$-definite pair-states 
   $\Upsilon (5S)^{(N/E)}$; Components of Initial Mixed-State $\Upsilon (5S)$. }
   \label{fig12}
 \end{figure}
In Figure {\ref{fig11}}, in the former-half, 
Experimental Data{\cite{Ref10,Ref11}} and 
Properties on Two-$h_{b}$'s and Two-$Z_{b}^{\pm}$'s 
are summarized with Problems
- {\it three-experimental facts}(Exp. F-1, F-2, F-3)- to 
be clarified on the standpoint of 
conventional Non-Relativ. classification: while, in the latter-half, 
Identification and 
Theoretical Interpretation in Evolved Classif. Scheme are given. 
Furthermore, here is also 
given the extended $CPT$-theorem, to be applied for Quantized Comp. Hadrons, which 
becomes a theoretical basis to solve the above problems. 

In Figure {\ref{fig12}}, we have shown the status of chirality conservation through all the 
relevant process in a rather concrete way. At the beginning it is to be stressed 
that the initial state $\Upsilon (5S)$ is produced through 
Electro-magnetic Process, ignorant of the chirality, 
and is the mixed-state of quantized definite-$\chi$ states, 
$\Upsilon (5S)^{(N)}$ and 
$\Upsilon (5S)^{(E)}$, with equal mixing-probability. 
Then{\cite{Ref12}} the status 
of $\chi_{b}$ 
conservation is described along the $\chi_{b}(+)$``Light''-chart and the 
$\chi_{b}(-)$``Shadow''-chart, respectively. 
On the basis of  all above considerations the problems 
in the Non-Relativistic scheme 
seem to be dissolved in Evolved Classif. Scheme, 
as follows: Exp. Fact-1 and -2 
on production rate are rather natural, and Exp. Fact-3 is predicted in 
$\widetilde{U}(4)_{DS,\mathfrak{m}}$-spin scheme. The problem on Spin-Flip 
Mech. disappears; since in our identification both of 
$\Upsilon^{(N)} (nS)$/($\Upsilon (nS)$) and 
$A^{(E)}(nS)$/($h_{b}\left(\left(1,2\right)P\right)$) is in the spin-triplet state. 
\section{Concluding Discussions}
\label{sec5}
\noindent
(Summary)

In this talk we have described essential logical basis of renewed covariant scheme of hadron-spectroscopy, developed recently, by applying propertime quantum mechanics to the multi-body confined quark system. 

We have applied it to investigate the phenomenology on the bottomonium and its alike systems, of which experiment has obtained a great progress recently. 

\noindent
(Remarks)

The results of above application seem us to be successful, and suggesting us that the phenomenological knowledge thus far obtained, resorting on non-relativistic scheme, should be reexamined from the relativistic one seriously and vigorously!
\subsubsection*{Acknowledgement}
I wish to express my deep gratitude to Prof. K. Yamada and Dr. T. Maeda for 
useful discussions and comments. 
This talk, without their kind cooperation, would not be realized. 
I am also grateful to Prof. T. Takamatsu, whose continual interests in Extension of 
Non-Relativistic to Covariant Spectroscopy, stimulating our collaboration in 
the $\sigma$-group. 

\end{document}